\documentclass{llncs}
\usepackage{mathtools}
\usepackage{amssymb}
\usepackage{amsmath}
\usepackage{amsfonts}
\usepackage{algorithm}
\usepackage{xcolor}
\usepackage{algpseudocode}
\usepackage{todonotes}
\makeatletter
\def\BState{\State\hskip-\ALG@thistlm}
\makeatother
\usepackage{multirow}
\usepackage{booktabs}
\usepackage{adjustbox}
\usepackage{graphicx}
\usepackage{subfigure}

\begin{document}

\title{An Active Learning Approach to the Falsification of Black Box Cyber-Physical Systems
}
\titlerunning{Active Learning Falsification for Black Box CPS}  
%
\author{Simone Silvetti\inst{1,2}
\and Alberto Policriti\inst{2,3} \and  Luca Bortolussi\inst{4,5,6} }
%
%
\tocauthor{Simone Silvetti, Alberto Policriti, and Luca Bortolussi }
\institute{
Esteco SpA, Trieste, Italy\\
\email{silvetti@esteco.com}\\
\and
Dima, University of Udine,  Udine, Italy\\
\email{alberto.policriti@uniud.it}\\
\and
Istituto di Genomica Applicata, Udine, Italy\\
\and
DMG, University of Trieste, Trieste, Italy\\
\email{luca@dmi.units.it}\\
\and
Modelling and Simulation Group, Saarland University, Saarbr\"ucken, Germany\\
\and
CNR-ISTI, Pisa, Italy}
\maketitle              

\begin{abstract}
Search-based testing is widely used to find bugs in  models of complex  Cyber-Physical Systems.
Latest research efforts have improved this approach by casting it as a falsification procedure of formally specified temporal properties, exploiting the robustness semantics of Signal Temporal Logic. 
The scaling of this approach to highly complex engineering systems requires efficient falsification procedures, which should be applicable also to black box models. 
Falsification
is also exacerbated by the fact that inputs are often time-dependent functions. 
We tackle the falsification of formal properties of complex black box models of Cyber-Physical Systems, leveraging  machine learning techniques from the area of Active Learning. Tailoring these techniques to the falsification problem with time-dependent, functional inputs, we show a considerable gain in computational effort, by reducing the number of model simulations needed. The goodness of the proposed approach is discussed on a challenging industrial-level benchmark from automotive.
\end{abstract}

\section{Introduction}
Model Based Development (MBD) is a well known design framework of complex engineered systems,  concerned with reducing cost and time of the prototyping process. Most prominently, this framework has been adopted in the industrial fields such as automotive and aerospace where the conformity of the end product is extremely important. The majority of  systems  in these areas are  Cyber-Physical Systems (CPS) \cite{baheti2011cyber}, where physical and software components interact producing complex behaviors. These systems can be described by appropriate mathematical models which are able to mime all the system behaviors. Moreover, it is necessary to have a suitable specification framework capable of analyzing the output of such models.

Hybrid Systems \cite{maler1991prom} are the mathematical framework usually adopted, while Temporal Logic \cite{pnueli1977temporal}, due to its ability to describe temporal events, is generally used as specification framework. The large expressivity of Hybrid Systems, which is the main reason of their success, is also the cause of their undecidability, even for simple logic formulas. Subsets of Hybrid Systems which are decidable for specific temporal logic formulas exist and have been widely studied during the last 15 years, as well as model checking techniques capable of verifying them \cite{alur1995algorithmic}. Unfortunately, the majority of CPS used nowadays in the industrial field are much more complex than decidable hybrid systems. They are mainly described by using block diagram tools (i.e Simulink/Stateflow, Scade, LabVIEW, and so on) where several switch blocks, 2/3-D look-up tables and state transitions coexist. These systems are generally not decidable and standard model checking techniques are not feasible, leading to the proposal of different techniques \cite{annpureddy2011s}.

Testing procedures with the purpose of verifying the model on specific behaviors have been adopted for several years. These are feasible approaches if it is possible in advance to write test cases which extensively cover all the possible events leading to system failure \cite{vinnakota1998analog}. With the increase of complexity, such an \emph{a priori} assumption is not feasible in most of the real cases and for this reason different techniques, such as random testing and search-based testing, have been introduced \cite{zhao2003generating}. The general idea consists in expressing the falsification procedure as an optimization process aimed to minimize a target quantity which describes how much a given property is verified. For example, achieving a negative value of the robustness semantics of a given Signal Temporal Logic (STL) \cite{donze2010robust} formula means falsifying the system with respect to that formula.

In this paper we study the falsification problem of black box systems (i.e block diagram models such as Simulink/Stateflow model or sets of ordinary differential equations generally used in automotive or aerospace industrial fields) which takes as input and produce as output continuous or Piecewise-Continuous (PWC) signals. The requirements are expressed by using STL. 

Solving such falsification problems in a search-based framework poses two main challenges. Generally, the simulation of block diagram models is time consuming, hence it is necessary to falsify the model with as few simulations as possible.
Moreover, the models accept continuous/PWC signals as inputs and an efficient finite dimensional parametrization is necessary to perform an optimization procedure. The contribution we propose in this paper is to tackle these challenges by a novel strategy leveraging Machine Learning techniques (Gaussian Processes and active learning) and by using a new adaptive version of the Control Points Parameterization approach. 

The paper is organized as follows. In Section \ref{sec:background} we review the definition of Dynamical System, Signal Temporal Logic and Gaussian Processes. In Section \ref{sec:domainEstimation} we discuss the Domain Estimation Problem which is solved by using Gaussian Processes. Section \ref{sec:falsification} presents the Falsification Approach and the adaptive optimization strategy performed by using the Gaussian Processes and adaptive function parameterization. In Section \ref{sec:probabilistic} we introduce the Probabilistic Approximation Semantics. In Section \ref{sec:tests} we briefly introduce the test cases and discuss the results. Finally in Section \ref{sec:conclusions} we provide the conclusions and discuss the future works.

\section{Background}\label{sec:background}
\subsection{Dynamical System}
We consider a system as a tuple $\mathcal{M}=(\mathcal{X},\mathcal{U},\mathtt{sim})$, where $\mathcal{X}$ and $\mathcal{U}$ are finite (or infinite) sets representing respectively the state of the system (coinciding for us with the output) and the input values. The system is equipped with a simulator, $\mathtt{sim}$, which is capable of computing the next state of the system starting from the current state and the input values. 
The simulator will be considered as a blackbox (i.e we can provide any input to the system and read the generated outputs).
The input set is $\mathcal{U}=V_0\times \dots \times V_n \times W_1\times \dots W_m$ where $V_i$ are finite sets and $W_i$ are compact sets in $\mathbb{R}$, representing respectively the discrete input events and the continuous input signals.
  The dynamics of the system is described with two functions: the \emph{state function} $\mathbf{x}:\mathcal{T} \to \mathcal{X}$ and the \emph{input function} $\mathbf{u}:\mathcal{T} \to \mathcal{U}$ which map each time $t \in \mathcal{T}$ to a state ($\mathbf{x}(t) \in \mathcal{X}$) and input ($\mathbf{u}(t) \in \mathcal{U}$), and where $\mathcal{T} =[0,T] \subset \mathbb{R}$.
  We call \textit{$k-th$ input signal} the $u_k$ function belonging to the input function $\mathbf{u}$ and identify with $\{\mathcal{T} \to \mathcal{U}\}$ the set of function from $\mathcal{T}$ to $\mathcal{U}$.

Given the input function $\mathbf{u}$ and the initial state $\mathbf{x}(0) = x_0$, the simulator determines the value of the state function on a discrete grid of time points from $\mathcal{T}$ according to:  
\begin{equation}\label{eq:sim}
\mathbf{x}(t_{k+1}) = \mathtt{sim}(\mathbf{x}(t_{k}),\mathbf{u}(t_{k}),h_k) 
\end{equation}
where $t_{k+1} - t_{k} = h_k>0 \in \mathbb{R}$ is the time-step used by the simulator to compute the dynamics of the system $\mathcal{M}$. 
A path or trajectory of the system $\mathcal{M}$ is 
\begin{equation}\label{eq:path}
\mathtt{Path}^\mathcal{M} =( (t_0,\mathbf{x}(t_0),\mathbf{u}(t_0)), \dots,(t_n,\mathbf{x}(t_n),\mathbf{u}(t_n)),\dots )
\end{equation}
which satisfies equation (\ref{eq:sim}).\\
In the engineering fields (such as Aerospace, Automotive \cite{weeks1995automotive}) the systems are usually described by means of block diagram tools, such as Simulink/StateFlow, $\dots$), each block representing a physical behaviour described by mathematical formulas (algebraic or differential equations) or by an external CAE software. The high complexity of these models justifies the blackbox assumption which we adopt.

\subsection{Signal Temporal Logic}
Signal Temporal Logic (STL,~\cite{maler2004monitoring}) is  a discrete linear time temporal logic used to reason about the future evolution of a path in continuous time. Generally this formalism is used to qualitatively describe the behaviors of trajectories of differential equations or stochastic models. The temporal operators we consider are all time-bounded and this implies that time-bounded trajectories are sufficient to assess the truth of every formula. The atomic predicates of STL are inequalities on a set of real-valued variables, i.e. of the form $\mu(\vec{X}) {:=} [g(\vec{X})\geq 0]$, where $g:\mathbb{R}^n \to \mathbb{R}$ is a continuous function and consequently  $\mu: \mathbb{R}^n \to \{\top,\bot\}$.

\begin{definition}
	\label{def:STL}
A formula $\phi\in\mathcal{F}$ of STL is defined by the following syntax:
\begin{equation}\label{eq:grammar}
	\phi := \bot \,|\,\top \,|\, \mu \,|\, \neg \phi \,|\,  \phi \vee \phi \,|\, \phi \mathbf{U}_{[T_1,T_2]} \phi, 
	\end{equation}
	where $\mu$ are atomic predicates as defined above, and $T_1<T_2<+\infty$.
\end{definition}
Eventually and globally modal operators can be defined as customary as $\mathbf{F}_{[T_1,T_2]}\phi \equiv \top \mathbf{U}_{[T_1,T_2]} \phi$ and $\mathbf{G}_{[T_1,T_2]}\phi \equiv \neg \mathbf{F}_{[T_1,T_2]}\neg \phi$.
STL formulae are interpreted over the dynamics $Path^{\mathcal{M}}$ of the model $\mathcal{M}$. We will consider the quantitative semantic \cite{donze2010robust} which, given a trajectory $\vec{x}(t)$, returns a real value capturing a notion of robustness of satisfaction whose sign captures the truth value of the formula (positive if and only if true), and whose absolute value gives a measure on how robust is the satisfaction.

\begin{definition}[Quantitative Semantics]
\label{def:pctmc_parameters}
	The quantitative satisfaction function $\rho:\mathcal{F}\times Path^{\mathcal{M}}\times [0,\infty)\rightarrow \mathbb{R}$ is defined by:
	\begin{itemize}
		\setlength\itemsep{-1em}
		\item $\rho(\top,\vec{x},t)= +\infty$\\
		\item $\rho(\mu,\vec{x},t) = g(\vec{x}(t))  \mbox{ where $g$ is such that } \mu(\vec{X}) \equiv [g(\vec{X}) \ge 0]$ \\
		\item $\rho(\neg \phi ,\vec{x},t) = - \rho(\phi,\vec{x},t)  $\\
		\item $\rho(\phi_1 \vee \phi_2,\vec{x},t) =\max( \rho(\phi_1,\vec{x},t), \rho( \phi_2,\vec{x},t)) $\\
		\item  $ \rho(\phi_1 \mathbf{U}_{[T_1,T_2]} \phi_2,\vec{x},t) = \underset{t'\in[ t+T_1, t+T_2]}{\sup} (\min(\rho(\phi_2,\vec{x},t), \underset{t''\in [t,t')}{\inf}  \rho(\phi_1,\vec{x},t) )) $
	\end{itemize}
\end{definition}

\subsection{Gaussian Processes}
\label{sec:gp}
Gaussian processes (GPs) are probabilistic methods used for classification or regression purposes. More specifically, a GP is a collection of random variables $X(t) \in \mathbb{R}$ ($t \in T$, an interval of $\mathbb{R}$) of which any finite number define a multivariate normal distribution. A GP is uniquely defined by its mean and covariance functions denoted respectively with $m: T \to \mathbb{R} $ and $k: \mathbb{R} \times \mathbb{R} \to \mathbb{R}$ such that for every finite set of points $(t_1,t_2,\dots,t_n)$: 
\begin{equation}
X \sim \mathcal{GP}(m,k) \iff (X(t_1),X(t_2),\dots,X(t_n))\sim \mathcal{N}(\mathbf{m},K)
\end{equation}
where $\mathbf{m} = (m(t_1),m(t_2),\dots,m(t_n))$ is the vector mean and $K \in \mathbb{R}^{n \times n}$ is the covariance matrix, such that $K_{ij} = k(X(t_i),X(t_j))$. From a functional point of view, GP is a probability distribution on the set of functions $X:\mathcal{T} \to R$. The choice of the covariance function is important from a modeling perspective because it determines the type of function that will be sampled with higher probability from a GP, see \cite{Rasmussen06gaussianprocesses}. In this work we use the Neural Network kernel, which performed better than more classical choices, like  Gaussian Radial Basis Function kernels, see \cite{Rasmussen06gaussianprocesses} for further details.

GPs are successfully used to solve regression problems starting from a training set with noisy observations,  $$((t_1,x_1),(t_2,x_2),\dots,(t_n,x_n))$$ 
The goal is to find a function $x:T \to \mathbb{R}$
such that $\forall i \le n, \quad x_i = x(t_i) + \epsilon$, and $\epsilon \sim \mathcal{N}(0,\sigma_n)$ (a gaussian noise is a common choice for regression with real-valued outputs).
In the GP paradigm a family of mean functions $m(x;h_1):\mathbb{R} \times H_1 \to \mathbb{R}$ and of covariance functions $k(x_1,x_2;h_2): \mathbb{R} \times \mathbb{R} \times H_2 \to \mathbb{R}$, where $h  = (h_1,h_2)$ are called hyperparameters,  are considered. The idea is to estimate the best hyperparameters which justify as much as possible, the observations provided in the training set. Mathematically it means to maximize the log marginal likelihood
$\max_{h} log\,p(x|t;h)$.
After having solved the previous optimization problem it is possible to predict the probability distribution of a new point as
$x(t^*) \sim \mathcal{N}(m^*,k^*)$,
where
$$m^* = (k(t^∗,t_1), \dots , k(t^∗,t_N))\hat{K}^{-1}_N x$$ 
$$k^*=k(t^∗,t^∗)−(k(t^∗,t_1), \dots , k(t^∗,t_N)) K^{-1}_N (k(t∗,t_1), \dots , k(t∗,t_N))^{T}$$

\section{Domain Estimation with Gaussian Processes}
\label{sec:domainEstimation}

\begin{definition}
	\label{def:domainEstimationProblem}
	Consider a function $f:\mathcal{D} \to \mathbb{R}$ and  an interval  $I\subseteq \mathbb{R}$. We define the \emph{domain estimation problem} requires us to identify a set $\mathcal{B}$ of size $|\mathcal{B}|=n$ of points  $x\in \mathcal{D}$ such that $f(x)\in I$: 
	\begin{equation}
		\mathcal{B}=\{ x \in \mathcal{D} | f(x) \in I \} \subseteq \mathcal{D},
	\end{equation}
\end{definition}
Gaussian Processes (GP) can be efficiently used to solve this task. Similarly with the Cross Entropy methods for optimization \cite{rubinstein2013cross}, the idea is to implement an iterative sample strategy in order to increase the probability to sample a point in $\mathcal{B}$, as the number of iterations increases.
Consider the the set $K(f) = \{(x_i,f(x_i)\,)\}_{i\le n}$ representing the partial knowledge we have collected after $n$ iterations and the GP $f_K(x) \sim GP(m_K(x),\sigma_K(x))$  trained on $K(f)$. We can easily estimate $P(x \in \mathcal{B}) = P(f_{K}(x) \in I)$ by computing the probability of a Gaussian distribution with mean $m_K(x)$ and variance $\sigma^2_K(x)$.
This corresponds to our uncertainty on the value of $f(x)$ belonging to $I$, as captured by the GP reconstruction of $f$. 
The previous probability can be effectively used to solve the domain estimation problem described in Definition \ref{def:domainEstimationProblem}. Our approach is described in Algorithm \ref{alg:domain}: 
\begin{itemize}
\item At initialization (line 2), we set the iteration counter ($i$) and the minimum distance ($d$) from the interval $I$. The set ($B$) containing the element of ($\mathcal{B}$) is set to empty, which ensures the algorithm is run at least once. The knowledge set $K(f)$ is initialized with some randomized points sampled from $\mathcal{D}$ (line 3).

\item In the iterative loop, the algorithm first checks if the number of counterexamples ($ce$) or if the maximum number of iterations ($maxIter$)  has been reached.
In this case, the method stops returning the estimated set ($B$) and the minimum distance from $I$ that has been registered until that point.
Otherwise new GPs are trained by using $K(f)$ (line 5) and a set composed by $m$ points ($D_{grid}$) is generated by Latin Hypercube sampling \cite{mckay1979comparison},  so to have an homogeneous distribution of points in space (line 6). For each of these points $x$, the probability $P(x \in \mathcal{B})=P(f_K(x) \in I)$ is calculated and the set $\{ (x, P(x \in \mathcal{B})), x \in  D_{grid} \}$ is created. Afterwards, a candidate point $x_{new}$ is sampled from $D_{grid}$  proportionally to its associated probability (line 7) so to increase the sampling of points with higher estimated probability of belonging to $\mathcal{B}$.
Consequently, $K(f)$ is upgraded and if $x \in \mathcal{B}$ then $x$ is added to $B$ (line 12). The procedure outputs also $d$,  the minimum distance of the evaluated points from the interval $I$ calculated during the procedure.
\end{itemize}

\begin{algorithm}[ht]
\caption{}\label{alg:domain}
\begin{algorithmic}[1]
\Procedure{[$B, d$ ] = domainEstimation($maxIter,ce,m,f,I$)}{}
\State $i \gets 0, \: B \gets \emptyset, \: d \gets +\infty$ 
\State $\textsc{initialize}(K(f))$
\While{( $ |B| \le ce \mbox{\textbf{ and }} i \le \mbox{maxIter})$}
    \State $f_{K(f)} \sim \textsc{trainGaussianProcess}(K(f))$
    \State $D_{grid} \gets \textsc{lhs}(m)$
    \State $x_{new} \gets \textsc{sample}\{ (x, P(x \in \mathcal{B})), x \in  D_{grid} \}$
\State $f_{new} \gets f(x_{new})$
\State $d \gets \min(d,\textsc{distance}(f_{new}, I))$
\State $K(f) \gets K(f) \cup \{(x_{new},f_{new})\} $
     \If{$f_{new} \in I$}
     \State $B=B \cup \{ x_{new} \}$   
    \EndIf
    \State $i \gets  i+1$
\EndWhile
\EndProcedure
\end{algorithmic}
\end{algorithm}

\section{The Falsification Process} 
\label{sec:falsification}

A big effort during the prototyping process consists in verifying the requirements usually expressed as safety property, such as:
\begin{equation}\label{eq:safety}
\forall (\mathbf{u},x_0) \in \{\mathcal{T} \to \mathcal{U}\}\times \mathcal{X}_0  \,\,, \rho(\phi,(\mathbf{u},\mathbf{x}),0)>0 
\end{equation}
meaning that for each input functions and initial state $x_0 \in \mathcal{X}_0 \subseteq \mathcal{X}$, the dynamics ($\mathtt{Path}^\mathcal{M}=(\mathbf{u},\mathbf{x})$) satisfies the STL formula $\phi$. 
It is possible to interpret the safety condition (\ref{eq:safety}) as a domain estimation problem associated with 
\begin{equation}
\label{eq:domainSet}
\mathcal{B}=\{(\mathbf{u},x_0) \in \{\mathcal{T} \to \mathcal{U}\}\times \mathcal{X}_0  \,\,, \rho(\phi,(\mathbf{u},\mathbf{x}),0)<0  \}
\end{equation}
with the purpose of verifying its emptiness, which entails that (\ref{eq:safety}) is satisfied. We call $\mathcal{B}$ the counterexample set and its elements counterexamples.

Solving the previous domain estimation problem could be extremely difficult because of the infinite dimensionality of the input space, which is a space of functions. For this reason, it is mandatory to parameterize the input function by means of an appropriate finite dimensional representation. 
One of the most used parameterization, due to its simplicity, is the \emph{fixed control point parameterization} (fixCP),
meaning that after having fixed the times $(t_0^k,\dots,t_{n_k}^k)$ the control points $\{(t_0^k, u_0^k),\dots,(t_{n_k}^k,u_{n_k}^k)\}$ are chosen as parameter of the $k$-th input signals. After having chosen an interpolation set of function with $n_k$ degrees of freedom for each $k$-th input signals ($\mathcal{P}_{n_k}^k \subset \{ \mathcal{T} \to \mathcal{U}_k\}$, e.g. piecewise linear, polynomials of degree $n_k$, and so on (see.\cite{sankaranarayanan2012falsification})), the fixCP parameterization will associate with each control point $c_k = \{(t_0^k, u_0^k),\dots,(t_{n_k}^k,u_{n_k}^k)\}$ the unique function $P_{c_k} \in \mathcal{P}_{n_k}^k$ satisfying $\forall i \le n\,,  P_{c_k}(t_i^k) = u_i^k$. 
Let us write $\mathcal{P}_\mathbf{n} = (\mathcal{P}_{n_0}^0, \dots,\mathcal{P}_{n_{|\mathcal{U}|}}^{|\mathcal{U}|})$, the set of interpolating functions.

It is clear that by increasing the number of control points, we will enlarge the set of  approximant functions $\mathcal{P}_\mathbf{n}$:  $\mathbf{n}\le \mathbf{m}$ implies $\mathcal{P}_{\mathbf{n}}  \subset \mathcal{P}_{\mathbf{m}}$, where $\mathbf{n}\le \mathbf{m}$  is intended pointwise. As piecewise linear or polynomial functions are known to be dense in the space of continuous functions, by choosing an appropriately large $\mathbf{n}$, we can approximate any input function with arbitrary precision.

Considering an \textbf{n}-fixCP it is possible to introduce the domain estimation problem \eqref{eq:domainSet} associated to the following set: 
\begin{equation}
\label{eq:domainSetFinite}
\mathcal{B}=\{(\bar{c},x_0) \in \mathcal{U}_1^{n_1}\times \dots \times  \mathcal{U}_{|\mathcal{U}|}^{n_{|\mathcal{U}|}} \times \mathcal{X}_0, \quad \rho (\phi,(P_\mathbf{n}(\bar{c}),x),0)<0  \}
\end{equation}
which , differently from \eqref{eq:domainSet}, is a finite dimensional set described by using $\sum_{j=1}^{|\mathcal{U}|} n_j + |\mathcal{X}_0|$ variables.

By the density argument it is clear that 
\begin{equation*}
\eqref{eq:domainSet} \mbox{ has at least one element }  \iff \exists \mathbf{n} \in \omega^{|\mathcal{U}|},\, \eqref{eq:domainSetFinite} \mbox{ has at least one element} 
\end{equation*}

A possible strategy is to solve the domain estimation problem associated with \eqref{eq:domainSetFinite} by choosing the minimum \textbf{n} such that $\mathcal{P}_\mathbf{n} \times X_0$ contains a counterexample. Applying that strategy, even in simple cases, could be cumbersome as shown in the following example.
\\

\noindent\textbf{Toy example}. Consider a simple  black box model which accepts a single piece-wise-constant functions $u:[0,1] \to [0,1]$ as input function, no initial state, and returning the same function $x=u$ as output. Considering the following requirement 
$\phi:= \neg (\textbf{G}_{[0,0.51]}\, 0<x<0.2 \wedge \textbf{G}_{[0.55,1]}\, 0.8 <x<1)$, it is evident that it could be falsified only in a CP having at least one control points $(t_i,u_i)$ such that $t_i \in [0.51,0.55)$. The minimum number of uniformed fixed control points necessary to achieve it is 9, which entails a considerable computational effort.

A natural way to overcome the limitation of the fixCP consists in considering the times of the control points as variables. 
A \emph{$\mathbf{n}$-adaptive Control Points parameterization} ($\mathbf{n}$-adaCPP) consists in a function $\bar{P}_{n_k}^k:\mathcal{T}^{n_k} \times \mathcal{U}_{k}^{n_k} \to \mathcal{P}_{n_k}^k$, which has twice as much  parameters than the fixed version: values at control points and times (which are constrained by $ \forall i < n\,\, t_i\le t_{i+1}$). 
The adaptive parameterization is preferable with respect to the fixed one because of its ability to describe functions with local high variability even with a low number of control points. In fact it is possible to concentrate any fraction of the available control points in a small time region, inducing a large variation in this region, while letting the parameterized function vary much less outside it. 

\subsection{Adaptive Optimization}
The idea of the adaptive optimization approach consists in falsifying \eqref{eq:safety} starting from a simple input function and increasing its expressiveness by increasing the number of control points.
Consider a model with input function $\mathbf{u}$ taking values in $\mathcal{U}_1 \times \dots \times \mathcal{U}_m$ and with initial state $x_0$ taking values in a closed set $\mathcal{X}_0 \subset \mathbb{R}^k$. After having defined a parameterization for each of the $m$ input signals,   Algorithm \ref{alg:ada} works as following:

\begin{itemize}
\item At the first iteration a parameterization $P_{\mathbf{n_0}} = \{P_1^0,\dots,P_n^0\}$ with zero control points for each signals ($\mathbf{n_0}=(0,\dots,0)$) is considered (line 2). Zero control points means defining input signals which are constant functions. The final counterexample set ($\Theta_f$) is set to empty, which ensures the optimization is run at least once (line 3).

\item In the iterative loop, the algorithm first checks if the number of counterexamples ($ce$) or if the maximum global number of iterations ($mgi$)  has been reached.
In this case, the method stops returning the counter example set ($B$). Otherwise,  the falsification problem is solved by using the domain estimation procedure \textsc{domainEstimation} (Algorithm \ref{alg:domain}) which returns the counterexample set and the minimum value of the robustness found by using that parameterization (see Section \ref{sec:domainEstimation} for details). The parameterization is then expanded by picking a coordinate of the input signal (line 6 - 10) and adding a new control point (line 11), obtaining a new parameterization $P_{\mathbf{n_{i+1}}}$.
\end{itemize}

\begin{algorithm}[!ht]
\caption{}\label{alg:ada}
\begin{algorithmic}[1]
\Procedure{[$B,d$] = adaptiveGPFalsification($mgi, mii, ce, m, \phi$)}{}
\State $\mathbf{n_0} \gets (0,\dots,0)$
\State $B \gets \emptyset,	\:k_0 \gets 0,\: i \gets 0,\: d_0 \gets +\infty$
\While{$ ( |B|\le ce \mbox{\textbf{ and }} i \le mgi)$}
    \State $[B^{-}, d_{i+1}] = \textsc{domainEstimation}(mii,\mathbf{n}_i,ce-|B|,m,\rho(\phi,\cdot,t),(-\infty,0))$
    \If{$d_{i+1} > d_{i}$}
    \State $ k_{i+1} \gets k_i$
    \Else
    \State  $ k_{i+1} \gets (k_i + 1) \mod n  $
    \EndIf
    \State  $\mathbf{n}_{i+1} \gets \mathbf{n}_{i}+ \mathbf{e}_k$
    \State $i \gets i+1$
    \State $B \gets B \cup  B^-$
\EndWhile
\EndProcedure
\end{algorithmic}
\end{algorithm}

The general idea of this approach is to keep low the number of parameters by starting from constant signal and gradually increasing the number of control points of the input functions. In the adaptive control points parameterization,  adding a control point means adding two new degrees of freedom (one for the time and one for the value of the control point). This means, one one side, having more expressiveness and so more chances to falsify the system, but on the other side this  complicates the optimization process and increases the dimension of the search space and hence the minimum number of simulation required to solve it.  
For this reason it is convenient to add control points only where it is truly necessary. 

\section{Probabilistic Approximation Semantics}
\label{sec:probabilistic}
Gaussian Processes can be used to estimate the probability that a given input falsifies a system as described in Section \ref{sec:domainEstimation} and Section \ref{sec:falsification}. 
This fact offers the possibility to define an approximate semantics which generalizes the concept of probability of falsification that we can infer considering the knowledge of the system we have collected. The basic idea is to decompose an STL formula as a Boolean combination of temporal modalities,  propagating the probability of the temporal operators, estimated by GPs, through the Boolean structure. 
Formally, let $\mathcal{S}_0$ be the subset of STL containing only atomic propositions and temporal formulae of the form $ \phi_1 \mathbf{U}_{[T_1,T_2]} \phi_2$, $\mathbf{F}_{[T_1,T_2]} \phi$ and $\mathbf{G}_{[T_1,T_2]} \phi$, that cannot be equivalently written as Boolean combinations of simpler formulae. Furthermore, let $\mathcal{S}$ be the propositional logic built over the set of atomic propositions $\mathcal{S}_0$.\footnote{$\phi\in\mathcal{S}$ iff $\psi :=  \phi \,|\, \neg \psi \,|\,  \psi \vee \psi \,|\, \psi \wedge \psi$, with $\phi\in\mathcal{S}_0$.} 

For simplicity, let us denote with $\theta$ a parameter and describing the input function with $u_\theta$ and the initial condition with ${x_0}_\theta$. We write $x_\theta$ to indicate the path generated by the simulator, given as input $u_\theta$ and ${x_0}_\theta$, accordingly to equation (\ref{eq:sim}). 
We want to define an (approximate) semantics describing the probability that a path $x_\theta$ satisfies a given formula $\psi\in\mathcal{S}$ (without simulating it). 
The idea is to evaluate the quantitative semantics of the atomic formulae $\phi_j\in\mathcal{S}_0$ of $\psi$ on a finite collection of  parameters ($\Theta = \{\theta_i\}_{i \le n}$), then building GPs in order to estimate  the probability  that the quantitative semantics of each formula $\phi_j$ is higher than zero on a target parameter. This is a Domain Estimation Problem (Section \ref{sec:domainEstimation})), where the function is the robustness associated to the STL formula $\phi_j$ and the interval $I$ is $(0,+\infty)$. We then propagate this probability through the Boolean structure of $\psi$ according to the following:
\begin{definition}[Probabilistic Approximation Semantics of S]
\label{def:probabilisticSemantics}
The probabilistic approximation function $\gamma:\mathcal{S} \times Path^{\mathcal{M}}\times [0,\infty)\rightarrow [0,1]$ is defined by:
	\begin{itemize}
		\setlength\itemsep{-1em}
		\item $\gamma(\phi,\theta,t)= P(f_{K(\phi)}(\theta)>0)$\\
		\item $\gamma(\neg \psi ,\theta,t) = 1 - \gamma(\psi,\theta,t)  $\\
		\item $\gamma(\psi_1 \wedge \psi_2,\theta,t) = \gamma(\psi_1,\theta,t) * \gamma(\psi_2,\theta,t)$\\
        \item $\gamma(\psi_1 \vee \psi_2,\theta,t) = \gamma(\psi_1,\theta,t) + \gamma(\psi_2,\theta,t) - \gamma(\psi_1 \wedge \psi_2,\theta,t)$\\
	\end{itemize}
    where $K(\phi) = \{ \theta_i,\rho(\phi,\theta_i,t) \}_{i=1,..,n}$ is the partial knowledge of the satisfiability of $\phi\in\mathcal{S}_0$ that we have collected  performing $n$ simulations for parameters $(\theta_i)_{i=1,..,n}$.
$f_{K(\phi)}$ is the GP trained on $K(\phi)$, and P refers to its probability.  For simplicity we use $\gamma(\psi,\theta,t)$ to mean $\gamma(\psi,(u_\theta,x_\theta),t)$.
\end{definition}
In the previous definition, the probability $P(f_{K(\phi)}(\theta)>0)$ is easily computed, as $f_{K(\phi)}(\theta)$ is Gaussian distributed.

Including the Probabilist Approximation Semantics (PAS) in our falsification procedure (Algorithm \ref{alg:ada}) is straightforward. Given the formula we have to falsify, first we negate and decompose it so to identify the $\mathcal{S}$ formula associated with it. Then we pass all the basic $\mathcal{S}_0$ formulas to the  \textsc{domainEstimation} procedure (Algorithm \ref{alg:domain}) and train a GP for each of them (instead of considering a single function (Algorithm \ref{alg:domain}), line 5). Afterwards we calculate its probabilistic approximation semantics to drive the sampling strategy (Algorithm \ref{alg:domain}, line 7). The rest of the algorithm remains the same. 
\\

\noindent \textbf{Remark.}
Consider the STL formula $\phi = \mathbf{G}_{[0,30]}(v\le160 \wedge \omega \le 4500)$. This formula is not in $\mathcal{S}_0$, as it can be rewritten as $\mathbf{G}_{[0,30]}(v\le160) \wedge \mathbf{G}_{[0,30]}(\omega \le 4500)$. We could have defined the set $\mathcal{S}_0$ in different ways (e.g. by including also $\phi$), our choice corresponding to the finer decomposition of temporal formulae. Even if this leads to an increased computational cost (more GPs have to be trained), it also provides more flexibility and allows us to exploit the boolean structure in more detail, as discussed in the following example.
\begin{figure}[]
 	\centering
 	\subfigure[PAS example] 
 	{ \label{fig:example1} \includegraphics[width=6cm]{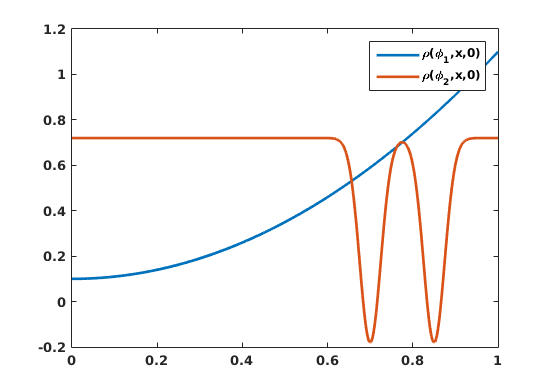}}
 	\hspace{2mm}
 	\subfigure[Automatic Transmission Req.] 
 	{ \label{fig:formulae} \includegraphics[width=5cm]{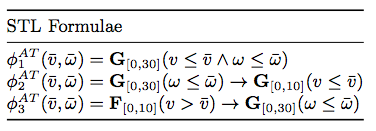}  }
 	\caption{(left) Example on the use of the probabilistic semantics. The curve treating the formula as a single one is the minimum of the two curves. (right) Requirements for the Automatic Transmission example of Section \ref{sec:tests}.
	}
 	\label{fig:example_and_table}
 \end{figure}
 
 \noindent \textbf{Example.} To clarify the advantages of the PAS, consider the functions $\rho(\phi_1,x,0)=x^2+1$ and $\rho(\phi_2,x,0)=-0.2+0.9(1-h(x,0.7,0.035)-h(x,0.85,0.035))$ representing  the robustness associated with the formulas $\phi_1$ and $\phi_2$ at time $0$ and for input parameter $x$, respectively. Here $h(x,m,s)$ is a gaussian function with mean $m$ and standard deviation $s$. We compare two approaches. In the first one, we calculate the probability of its negation i.e $\gamma( \neg (\phi_1 \wedge \phi_2),x,0) = 1 - \gamma( (\phi_1 \wedge \phi_2),x,0) $ by means of a single gaussian process. In the second one, we decompose the conjunction  and calculate its PAS  $ \gamma( \neg (\phi_1 \wedge \phi_2),x,0) =1 - \gamma( \phi_1,x,0)*\gamma( \phi_2,x,0) $ by means of two separated Gaussian Processes. 
Functions used by the method to drive the sample are represented in Figure \ref{fig:example1}. 
In the first case, the signal which is smooth in $[0,0.65]$ and highly variable between $(0.65,1]$, and the method to sample a lot of points near $x=0$, as the function  is close to zero near this point, hence a potential falsification point until the precision of the reconstruction is sufficiently high to discard this option. This requires $55.35 \pm 45.10$ function evaluations. On the contrary the second approach shows a rapid discovery of the falsification area i.e $17.19 \pm 7.71$ evaluations, because the two components are treated independently, and the method quickly finds the minima regions of  $\gamma( \phi_2,x,0)$, after an initial phase of homogeneous exploration. In addition,  the paraboloid $\gamma( \phi_1,x,0)$ is smooth and requires few evaluations for a precise reconstruction.

\section{Case Studies and Results} 
\label{sec:tests}
 
In this section we discuss a case studies to illustrate our approach, taken from \cite{sankaranarayanan2012falsification}. We will compare and discuss the performance of a prototype implementation in Matlab of our approach with S-Taliro toolbox \cite{fainekos2012verification}. We use S-Taliro to compute the robustness, and the implementation of Gaussian Process regression provided by Rasmussen and Williams \cite{rasmussen2010gaussian}.
\\

\noindent \textbf{Automatic Transmission (AT)}. We consider a Simulink model of a Car Automatic Gear Transmission Systems. There are two inputs: the throttle and the brake angle dynamics describing the driving style. Modes have two continuous state variables, describing vehicle ($v$) and engine ($\omega$) speed.  The Simulink model is initialized with a fixed initial state $(w_0,v_0)=(0,0)$, it contains 69 blocks (2 integrators, 3 look-up table, Stateflow Chart, \dots). The requirements are described by means of STL formulae as reported in Figure \ref{fig:formulae}. The first requirement ($\phi_1^{AT}$) is a so called \textit{invariant} which says that in the next 30 seconds the engine and vehicle speed never reach $\bar \omega$ rpm and $\bar{v}$ km/h, respectively. The second requirement ($\phi_2^{AT}$) says that if the engine speed is always less than $\bar \omega$ rpm, then the vehicle speed can not exceed $\bar{v}$ km/h in less than 10 sec.  
Finally, the third requirement ($\phi_3^{AT}$) basically says that within 10 sec. the vehicle speed is above $\bar{v}$ km/h and from that point on the engine speed is always less than $\bar \omega$ rpm.
\\

\begin{table}[!t]

	\centering
	\caption{Results. All the times are expressed in seconds. Legend -  nval: number of simulations, f: fraction of optimization run that falsify the property, Alg: the algorithm used as described in Section \ref{sec:tests}. }
	\label{my-label}
	\begin{adjustbox}{max width=\textwidth}
		\begin{tabular}{@{}c||cc|cc|ccc@{}}
			\toprule
			\multicolumn{1}{c}{}& \multicolumn{2}{c}{Adaptive PAS} & \multicolumn{2}{c}{Adaptive GP-UCB} & \multicolumn{3}{c}{S-TaLiRo} \\ \cmidrule{1-8}
			 Req &    nval    & times    &     nval    & times        & nval    & times    & Alg  \\ \midrule
			 $\phi_1^{AT}(160,4500)$ & $4.42\pm0.53$  & $ 2.16\pm0.61 $         & $4.16\pm2.40$  & $ 0.55\pm0.30 $                  &$5.16\pm4.32$         &$0.57\pm0.48$            & UR    \\
			 $\phi_1^{AT}(160,4765)$ & $6.90\pm2.22$       & $ 5.78\pm3.88 $          &  $8.7\pm1.78$  & $ 1.52\pm0.40 $                  & $39.64\pm44.49$        & $4.46\pm4.99$              & SA \\
			 $\phi_2^{AT}(75,4500)$ & $3.24\pm1.98$         &$1.57\pm1.91$          & $7.94\pm3.90$  & $1.55\pm1.23 $                  & $12.78\pm11.27$        & $1.46\pm1.28$              &CE \\ 
			 $\phi_2^{AT}(85,4500)$ & $10.14\pm2.95$         &$12.39\pm6.96$          & $23.9\pm7.39$  & $ 9.86\pm4.54 $                  & $59\pm42$        & $6.83\pm4.93$              &SA \\
			 $\phi_2^{AT}(75,4000)$ & $8.52\pm2.90$         &$9.13\pm5.90$          & $13.6\pm3.48$  & $ 4.12\pm1.67 $                  & $43.1\pm39.23$        & $4.89\pm4.43$              &SA \\
			 $\phi_3^{AT}(80,4500)$ & $5.02\pm0.97$         &$2.91\pm1.20$          & $5.44\pm3.14$  & $ 0.91\pm0.67 $                  & $10.04\pm7.30$        & $1.15\pm0.84$              &CE \\
			 $\phi_3^{AT}(90,4500)$ & $7.70\pm2.36$         &$7.07\pm3.87$          & $10.52\pm1.76$  & $ 2.43\pm0.92 $                  & $11\pm9.10$        & $1.25\pm1.03$              &UR \\
			 
			 \midrule
		\end{tabular}
	\end{adjustbox}
\end{table}

\noindent\textbf{Results.} We analyze the performance of our approach in terms of the minimum number of simulations and computational time needed to falsify the previous test cases. We have performed 50 optimization runs for each STL formula and compared its performance with the best statistics achieved among a Cross Entropy (CE), Montecarlo Sampling (SA) and Uniform Random Sampling (UR) approaches performed with the S-TaLiRo tool \cite{annpureddy2011s} and the GP-UCB algorithm applied to falsification as described in \cite{akazaki2016falsification}. As the table shows, we have good results in terms of the minimum number of evaluations  needed to falsify the systems with respect to the STL formulae, outperforming in almost all tests the methods of the S-TaLiRo suite and the GP-UCB approach. This is the most representative index, as in real industrial cases the simulations can be considerably expensive (i.e cases of real measurements on power bench, time and computation intensive simulations).   In these cases the total computational time is directly correlated with the number of simulations and the time consumed by the optimizer to achieve its strategy becomes marginal. Furthermore, we are testing our method with a prototype implementation which has not been optimized, in particular for what concerns the use of Gaussian Processes. Despite this, the numerical results in terms of minimum number of simulations are outperforming S-TaLiRo and GP-UCB approach.

\noindent \textbf{Conditional Safety Properties.}
When we define a conditional safety property i.e  $\mathbf{G}_{T}(\phi_{cond} \to \phi_{safe})$ we would like to explore  cases in which the the formula is falsified but the antecedent condition holds (see \cite{akazaki2016falsification}). This is particular relevant when the formula cannot be falsified, as it reduces the search space, ignoring regions where the formula is trivially true due to a false antecedent. Focusing on the region where  $\phi_{cond}$ holds requires a straightforward modification of the sampling routine of the Domain Estimation Algorithm (Algorithm \ref{alg:domain}, line 6-7). Instead of performing the sampling directly on the input provided by the Latin Hypercube Sampling Routine (Algorithm \ref{alg:domain}, line 6), we previously define a set of inputs verifying the antecedent condition (by the standard Domain Estimation Algorithm using the Gaussian Processes trained on the robustness of the antecedent condition) and then we sample from this set the candidate point (Algorithm \ref{alg:domain}, line 7).

To verify the effectiveness of this procedure, we calculate the percentage of sampled inputs satisfying the antecedent condition of the STL formula $\mathbf{G}_{[0,30]}( \omega \le 3000 \to v\le 100) $, which cannot be falsified. This percentage is   43\% for the GP-UCB algorithm, but increases to 87\% for the modified domain estimation algorithm.

\section{Conclusions}
\label{sec:conclusions}

In this paper we propose an adaptive strategy to find bugs in black box systems. We search in the space of possible input functions, suitably parameterized in order to make it finite dimensional. We use a separate parameterization for each different input signal, and we use an adaptive approach, increasing gradually the number of control points as the search algorithm progresses. This allows us to solve falsification problems of increasing complexity, looking first for simple functions and then for more and more complex ones. 
The falsification processes is then cast into the Domain Estimation Problem framework, which use the Gaussian Processes to constructs an approximate probabilistic semantics of  STL formulae, giving high probability to regions where the formula is falsified.
The advantage of using such an approach is that it leverages the Bayesian emulation providing a natural balance between exploration and exploitation, which are the key ingredients in a search-based falsification algorithm.
In addition to a novel use of Gaussian Processes, we also rely on a new  adaptive parameterization,  treating the time of each control point as a variable, thus leading to an increase in expressiveness and flexibility, as discussed in the Section \ref{sec:falsification}. Moreover with a slight modification of our algorithm we efficiently manage the falsification of the condition safety properties, increasing the efficiency of the usual GP-UCB algorithm in focussing the search on the region of points satisfying the antecedent. 
\\

\noindent \textbf{Future Work.} The experimental results are quite promising, particularly for what concerns  the  number of simulations required to falsify a property, which is lower than other approaches. The computational time of the current implementation, however, is in some cases higher then S-TaLiRo. The main problem is in the cost of computing predictions of the function emulated with a Gaussian Process (GP). This cost, in fact, is proportional to the number of already evaluated inputs used to train the GP. To reduce this cost, we can leverage the large literature about sparse representation techniques for GP \cite{Rasmussen06gaussianprocesses}.
Furthermore, with the increase in the number of control points, we face  a larger dimensional search space, reflecting in an increased number of simulations needed to obtain an accurate representation of the robustness for optimization, with a consequent increase of computational time.  We can partially ameliorate this problem, typical  of naive implementations of the Bayesian approach,  by refining the choice of the covariance function and/or constraining some of its hyperparameters so to increment the exploration propensity of the search. 
In the future,  we also plan  to improve the adaptive approach which is in charge of increasing the control points of an input signal, with the goal of dropping control points that are not useful.

In the current paper, we use the GP-based sampling scheme to deal efficiently with falsification. However, our approach can be modified to deal with the coverage problem \cite{dreossi2015efficient}, i.e. the identification  identification of a given number of counterexamples which are homogeneously distributed in the falsification domain. Our idea is to modify the sampling  algorithm (Algorithm. \ref{alg:domain}, line 7) by adding a memory of already visited areas, so to distribute samples homogeneously in space. 
\\

\noindent \textbf{Related Work.} Different approaches have been proposed to achieve the falsification of black box models, starting from test based approaches until recently, when search-based test approaches have become more popular. Stochastic local search \cite{deshmukh2015stochastic}, probabilistic Monte Carlo \cite{abbas2013probabilistic} and mixed coverage/guided strategy \cite{dreossi2015efficient} approaches have been proposed and benchmark problems  created \cite{jin2014powertrain,Hoxha2015}. Two software packages \cite{donze2010breach,annpureddy2011s} implement the aforementioned techniques.
Both these software tools assume a fix parameterization of the input function, differently from us. Similarly to our approach, in  \cite{dreossi2015efficient} and \cite{deshmukh2015stochastic}  the fixed parameterization is avoided. More specifically  in \cite{dreossi2015efficient} no parameterization has been used at all and the input signals are modified on the fly based on the robustness of the partial system trajectories. In \cite{deshmukh2015stochastic} a uniform discretization of the input domains (both time and values) is dynamically applied to discretize the search space. The use of Gaussian Processes for falsification has been adopted in \cite{akazaki2016falsification} but it is restricted to Conditional Safety Properties.    
\bibliography{biblio}{}
\bibliographystyle{plain}

\end{document}